\documentclass[prx,twocolumn,superscriptaddress,nopacs]{revtex4}
\usepackage{graphicx}

\usepackage{amsmath}
\usepackage{amssymb}
\usepackage{hyperref}
\usepackage{color}

\newcommand{\tot}{\mathrm{tot}}

\begin{document}

\title{Exact Ground State of Lieb-Mattis Hamiltonian as a Superposition of N\'{e}el states}
\author{Louk Rademaker}
\affiliation{Department of Theoretical Physics, University of Geneva, 1211 Geneva, Switzerland}

\date{\today}

\begin{abstract} 
We show that the exact ground state of the Lieb-Mattis Hamiltonian is an equal-weight superposition of all possible classical N\'{e}el states, and provide an exact formulation of this superposition in the $z$-spin basis for both $S=1/2$ and general $S$ using Schwinger bosons. In general, a superposition of possible rotations on a general initial state is symmetric if and only if the initial state has a nonzero overlap with a singlet state and is otherwise made up of states that vanish due to the symmetrization. Most notably, $|s, m=0 \rangle$ states will vanish if symmetrized, which explains how a superposition of N\'{e}el states projects onto its singlet component.
\end{abstract}

\maketitle

\section{Introduction}
The ground state of finite systems that exhibit spontaneous symmetry breaking in the thermodynamic limit is typically still symmetric and unique.
In particular, this has been proven for the Heisenberg antiferromagnet.\cite{Marshall:1955aa,Lieb:1962hn} To arrive at such proof of uniqueness, Lieb and Mattis considered only the $k=0,\pi$ part of the Heisenberg Hamiltonian, which is now known as the {\em Lieb-Mattis Hamiltonian}.

The interesting thing about the Lieb-Mattis Hamiltonian is that it displays the same singular behavior in the thermodynamic limit as is common for spontaneous symmetry breaking. That is, upon adding a symmetry breaking field $\hat{H}' = - B (\hat{S}^z_A - \hat{S}^z_B)$, the limits $B \rightarrow 0^+$ and $N \rightarrow \infty$ do not commute. If one keeps $B$ finite while taking the thermodynamic limit, the resulting ground state is the classical N\'{e}el state antiferromagnet.

Given that in the thermodynamic limit the N\'{e}el state and the exact symmetric ground state become degenerate, one might wonder what their relationship is. In particular, can we take a suitable superposition of all possible N\'{e}el states to construct the symmetric ground state? Here we show that the answer to this question is yes. In fact, an equal-weight superposition of all possible directions a N\'{e}el state is equal to the Lieb-Mattis ground state.

This led us to study more general the properties of symmetric superpositions of some initial polarized state. We show that, depending on the definition of your rotation operator, some states vanish while others are projected onto states of the form $|s,m=0\rangle$. Consequently, we conclude that the `superposition of all directions' is symmetric if and only if the initial state has a nonzero overlap with a singlet and is otherwise made up of states that vanish.

In the remainder of this manuscript we will first introduce rotations and the symmetrized superposition in Sec.~\ref{Sec:def}. In Sec.~\ref{Sec:Neel} we show, using three complementary approaches, that a superposition of N\'{e}el states is equal to the Lieb-Mattis ground state. In Sec.~\ref{Sec:Generalized} we derive the structure of a general superposition of directional states. Finally, in Sec.~\ref{Sec:Outlook} we provide some outlook of our results.

\section{Definitions}
\label{Sec:def}
A state $|\psi \rangle$ is invariant under $SU(2)$ rotations if and only if it is an eigenstate of every possible spin rotation. Note that due to the non-Abelian nature of $SU(2)$, this implies that only states in the trivial representation, that is, singlet states with $\hat{{\bf S}}_{\tot}^2 = 0$, are $SU(2)$ invariant.

Consider a state $| \psi_0 \rangle$ that is {\em not} $SU(2)$ invariant. Without loss of generality, we consider it to have a polarization in the $z$-direction. Similar states in different directions can be constructed as follows,
\begin{equation}
	| \theta, \phi \rangle = e^{-i \phi \hat{S}^z } e^{-i \theta \hat{S}^y} | \psi_0 \rangle
	\label{Eq:DirectionalStates}
\end{equation}
which we will call a {\em directional state}. We define the {\em symmetrized} state as the equal-weight superposition of these different directional states,
\begin{equation}
	| \psi_S \rangle = \int \; \sin \theta \; d\theta \; d\phi \; | \theta, \phi \rangle.
	\label{Eq:Symmetrized}
\end{equation}
The definition of Eq.~\eqref{Eq:DirectionalStates} is not unique. 
Though our definition is consistent with most of the literature (for example Ref.~\cite{Klauder:1985ta}), in some instances a different definition of the spin rotation is chosen. For example, Ref.~\cite{Gazeau:2009vb} defines directional states as
\begin{equation}
	| \theta, \phi \rangle_G  = e^{\xi \hat{S}^+ - \xi^* \hat{S}^-} | s,s \rangle
	\label{Eq:WrongDef}
\end{equation}
with $\xi = -\frac{\theta}{2} e^{-i \phi}$ and $|s,s\rangle$ the maximally polarized state with total spin $s$. These states differ a phase factor $e^{i s \phi}$ from our definition Eq.~\eqref{Eq:DirectionalStates}. However, as we shall see, for a superposition of N\'{e}el states these phase factors exactly cancel, and therefore Eq.~\eqref{Eq:WrongDef} and Eq.~\eqref{Eq:DirectionalStates} yield the same results. Only for the more general superposition discussed in Sec.~\ref{Sec:Generalized}, the precise definition of rotation matters.

Furthermore, observe that if one chooses the original not-invariant state to be the maximally polarized state $|s,s \rangle$, the definition Eq.~\eqref{Eq:DirectionalStates} defines spin coherent state. The symmetrized state of Eq.~\eqref{Eq:Symmetrized} is thus a superposition of all possible spin coherent states.

\section{Symmetrized N\'{e}el wavefunction}
\label{Sec:Neel}

In this section we will show that a symmetrized superposition over N\'{e}el states is equal to the ground state of the Lieb-Mattis Hamiltonian. For this, consider a spin $S$ system on a bipartite lattice. The N\'{e}el state polarized in the $z$ direction is a product state with $|S \rangle$ on all sites in sublattice $A$ and $| -S \rangle$ on all $B$ sites,
\begin{equation}
	| \psi_N \rangle = \prod_{i \in A} | S \rangle_i \; \prod_{j \in B} | -S\rangle_j.
\end{equation}
The Lieb-Mattis Hamiltonian is defined as
\begin{eqnarray}
	\hat{H}_{LM} &=& \frac{1}{N} \hat{\bf S}_A \cdot \hat{\bf S}_B 
		\label{eq:HLM} \\
	&=& \frac{1}{2N} \left( \hat{{\bf S}}_{\mathrm{tot}}^2 -  \hat{\bf S}_A^2 -  \hat{\bf S}_B^2 \right)
	\label{eq:HLM2}
\end{eqnarray}
For a bipartite lattice with $N$ sites and on each site a spin $S$, the unique ground state is a total singlet state ($S_\tot=0$) with maximal sublattice spin $S_A = S_B = NS/2$ and ground state energy $E_0 = -\frac{S}{2} (\frac{N S}{2} + 1)$.\cite{VanWezel:2008aa} In order to show that a state is the ground state, it suffices therefore to show that it has maximal sublattice spin and total spin zero.

\subsection{General structure}
The symmetrized superposition of N\'{e}el states in all possible directions is
\begin{equation}
	|\psi_S \rangle = \frac{1}{4\pi}
		\int_0^\pi \sin \theta \; d\theta \; \int_0^{2\pi} d\phi \;
		 e^{-i \hat{S}^z_\tot \phi} e^{-i \hat{S}^y_\tot \theta} | \psi_N \rangle,
	\label{Eq:SymmetrizedNeel}
\end{equation}
see also Sec. 4.2 of Ref~\cite{Tasaki:2019book}.

Our claim is that this state is the ground state of the Lieb-Mattis Hamiltonian. Since $\hat{\bf S}_{A/B}$ commute with the rotation operator $e^{-i \hat{S}^z_\tot \phi} e^{-i \hat{S}^y_\tot \theta}$, and $\hat{\bf S}_{A/B}^2  | \psi_N \rangle = (NS/2)(NS/2+1) $, it follows that $|\psi_S \rangle$ has maximal sublattice spin $S_{A/B} = NS/2$. The only remaining thing to prove is that $|\psi_S\rangle$ is nonzero \'{a}nd it has $S_\tot = 0$.

First, we use that fact that the N\'{e}el state can be obtained by projecting the Lieb-Mattis ground state $|\psi_{LM}\rangle$ onto the space with $S^z_A = NS/2$ and $S^z_B=-NS/2$,
\begin{eqnarray}
	| \psi_N \rangle 
	&=& \hat{P}_{S^z_A = NS/2} \hat{P}_{S^z_B = NS/2} | \psi_{LM} \rangle
\end{eqnarray}
It follows that $\langle \psi_{LM} | \psi_N \rangle \neq 0$. The overlap between the symmetrized wavefunction $| \psi_S \rangle$ and the Lieb-Mattis ground state is therefore nonzero too,
\begin{eqnarray}
	\langle \psi_{LM} | \psi_S \rangle & = &
		\frac{1}{4\pi}
		\int_0^\pi \sin \theta \; d\theta \; \int_0^{2\pi} d\phi \; \nonumber \\ && \;\;
		\langle \psi_{LM} |
		 e^{-i \hat{S}^z_\tot \phi} e^{-i \hat{S}^y_\tot \theta} | \psi_N \rangle \\
	& = &
		\frac{1}{4\pi}
		\int_0^\pi \sin \theta \; d\theta \; \int_0^{2\pi} d\phi \;
		\langle \psi_{LM} | \psi_N \rangle \\
	 & = &\langle \psi_{LM} | \psi_N \rangle \neq 0. 
\end{eqnarray}
This implies that $|\psi_S \rangle$ itself is non-vanishing.

To prove that $|\psi_S \rangle$ is a total spin singlet, we observe that the projection $\hat{P}_{S^z_A = NS/2} \hat{P}_{S^z_B = NS/2} $ is a function that only depends on the $z$-component of the spin, $\hat{S}^z_\tot$. Let us call this function $W(\hat{S}^z_1, \ldots , \hat{S}^z_N )$. Let $\hat{U}$ be an arbitrary uniform spin rotation, such that $\hat{U} \hat{S}^z_j \hat{U}^\dagger = {\bf n} \cdot \hat{\bf S}_j$ for all $j$. Since the Lieb-Mattis ground state satisfies $\hat{U} | \psi_{LM} \rangle = |\psi_{LM} \rangle$, we have
\begin{eqnarray}
	&&\hat{U} W (\hat{S}^z_1, \ldots , \hat{S}^z_N ) | \psi_{LM} \rangle 
		= \hat{U} W (\hat{S}^z_1, \ldots , \hat{S}^z_N )  \hat{U}^\dagger| \psi_{LM} \rangle 
	 \nonumber \\
	 && \phantom{m} = W ({\bf n} \cdot \hat{\bf S}_1, \ldots , {\bf n} \cdot \hat{\bf S}_N ) | \psi_{LM} \rangle
	 \equiv W_{\bf n} | \psi_{LM} \rangle
\end{eqnarray}
where $W_{\bf n}$ only depends on the direction ${\bf n }$ and not on the specific choice of rotation operator $\hat{U}$. It follows that specifically for the N\'{e}el state,\footnote{And other states that can be expressed as $W(\hat{S}^z_1, \ldots , \hat{S}^z_N ) | \psi_0 \rangle$ with $W$ some function and $|\psi_0 \rangle$ a singlet state.} the precise choice of rotation operator $\hat{U}$ is not relevant and we can express the symmetrized wavefunction as
\begin{equation}
	| \psi_S \rangle = \int_{|{\bf n}| = 1} \frac{d {\bf n}}{4\pi} \;
	 W_{\bf n} | \psi_{LM} \rangle
\end{equation}
which is manifestly $SU(2)$ invariant and therefore a singlet.

We have thus proven that $|\psi_S \rangle$ given by Eq.~\eqref{Eq:SymmetrizedNeel} is the ground state of the Lieb-Mattis Hamiltonian.\footnote{This section is based on private communications with H. Tasaki.}

\subsection{Explicit construction for $S=1/2$}

The previous paragraph contained an elegant and general proof that the symmetrized state $|\psi_S \rangle$ is the ground state of the Lieb-Mattis Hamiltonian. In this paragraph, we will construct explicitly this symmetrized wavefunction for $S=1/2$. Since any N\'{e}el state is a product state, we can write down the rotated N\'{e}el state as
\begin{equation}
	|\theta, \phi \rangle = \prod_{i \in A} |\psi_{iA}(\theta,\phi) \rangle
	 \prod_{j \in B} |\psi_{jB}(\theta,\phi) \rangle
	\label{eq:RotatedNeel1}
\end{equation}
where, for $S=1/2$,
\begin{equation}
	|\psi_{iA}(\theta,\phi) \rangle = e^{-i \phi/2} \cos \frac{\theta}{2} | \uparrow_i \rangle
		+ e^{i \phi/2} \sin \frac{\theta}{2} | \downarrow_i \rangle
	\label{eq:RotatedNeel2}
\end{equation}
and
\begin{equation}
	|\psi_{jB}(\theta,\phi) \rangle = - e^{-i \phi/2} \sin \frac{\theta}{2} | \uparrow_j \rangle
		+ e^{i \phi/2} \cos \frac{\theta}{2} | \downarrow_j \rangle.
	\label{eq:RotatedNeel3}
\end{equation}
Using this explicit expression, we can construct the symmetrized wavefunction $|\psi_S \rangle$ in the basis of spin configurations $\mathcal{C}$
\begin{equation}
	|\psi_S \rangle = \sum_{\mathcal{C}} a_{\mathcal{C}} | \left\{ \mathcal{C} \right\} \rangle
	\label{eq:ExpansionPsiS}
\end{equation}
in the given quantization axis (here chosen to be $z$). It follows directly from Eqns.~\eqref{eq:RotatedNeel1}-\eqref{eq:RotatedNeel3} that the amplitude $a_{\mathcal{C}}$ only depends on the number of $\uparrow,\downarrow$ spins on the $A/B$ sublattices,
\begin{eqnarray}
	a_{\mathcal{C}} &=& \int_0^\pi \sin \theta \; d\theta \; \int_0^{2\pi} d\phi \;
	\nonumber \\ && 
	 \left( e^{-i \phi/2} \cos \frac{\theta}{2} \right)^{N^\uparrow_A}
		\left( e^{i \phi/2} \sin \frac{\theta}{2}\right)^{N^\downarrow_A}
	\nonumber \\ && \times
		\left( - e^{-i \phi/2} \sin \frac{\theta}{2} \right)^{N^\uparrow_B}
		\left( e^{i \phi/2} \cos \frac{\theta}{2} \right)^{N^\downarrow_B}.
\end{eqnarray}
We will first do the integral over $\phi$. Because the number of sites $N$ is even, we notice that $-N^\uparrow_A+N^\downarrow_A-N^\uparrow_B+N^\downarrow_B = N^\uparrow - N^\downarrow$ is even as well. This implies that the integral over $\phi$ vanishes {\em unless} $N^\uparrow = N^\downarrow$. This means that $|\psi_S\rangle$ is an eigenstate of $\hat{S}^z_{\mathrm{tot}}$ with eigenvalue zero.

The $S^z_{\mathrm{tot}}=0$ condition allows us to write all parameters $N^{\uparrow/\downarrow}_{A/B}$ strictly as a function of $N_B^\uparrow$ and $N$,
\begin{eqnarray}
	N_A^\uparrow &=& N/2 - N_B^\uparrow, \\
	N_A^\downarrow &=& N_B^\uparrow, \\
	N_B^\downarrow &=& N_A^\uparrow=N/2 - N_B^\uparrow . \\
\end{eqnarray}
The remaining integral over $\theta$ is obtained by using the identity
\begin{equation}
	I_\theta = \int_0^\pi \sin \theta d\theta  
	\; \cos^k \frac{\theta}{2} \sin^m \frac{\theta}{2}
	= \frac{2 (k/2)! (m/2)!}{((k+m)/2+1)!}.
	\label{Eq:IntIdentity}
\end{equation}
Throwing out an overall $N_B^\uparrow$-independent prefactor, we find that the {\em unnormalized} amplitudes are
\begin{equation}
	a_{N_B^\uparrow} \propto \frac{(-1)^{N^\uparrow_B}}{\binom{N/2}{N_B^\uparrow}}.
	\label{eq:aPreNorm}
\end{equation}
We only need to establish the normalization. To do so, observe that for a system with $N$ spins, the number of states with $S^z_\tot = 0$ and a given $N^\uparrow_B$ is $\binom{N/2}{N_B^\uparrow}^2$.\footnote{One can verify that indeed the total number of states with $S^z_\tot=0$ is given by $\binom{N}{N/2} = \sum_{N_B^\uparrow = 0}^{N/2} \binom{N/2}{N_B^\uparrow}^2$.} This binomial precisely cancels the binomial in Eqn.~\eqref{eq:aPreNorm}, and because there are $N/2+1$ different possible sectors with fixed $N_B^\uparrow$, we conclude that the proper normalized amplitudes are
\begin{equation}
	a_{N_B^\uparrow}  = \frac{(-1)^{N^\uparrow_B}}{\sqrt{N/2+1} \binom{N/2}{N_B^\uparrow}}.
	\label{eq:Amplitudes}
\end{equation}
Note that the sign of the amplitude is completely determined by $N^\uparrow_B$ according to Marshall's sign rule for the singlet ground state of antiferromagnetic systems.\cite{Marshall:1955aa}

Next, we will show that the symmetric wavefunction $|\psi_S \rangle$ from Eqn.~\eqref{eq:ExpansionPsiS} with amplitudes Eqn.~\eqref{eq:Amplitudes} is the ground state of the Lieb-Mattis Hamiltonian. To do this, we explicitly write out Eqn.~\eqref{eq:HLM},
\begin{equation}
	\hat{H}_{LM} = \frac{1}{N} \sum_{i \in A, j \in B} \left( \hat{S}^z_i \hat{S}^z_j 
	+ \frac{1}{2} \left( \hat{S}^+_i \hat{S}^-_j + \hat{S}^-_i \hat{S}^+_j \right) \right)
\end{equation}
We then compute the amplitudes $a'_{\mathcal{C}}$ of each configuration $\mathcal{C}$ with given $N_B^\uparrow$ in the vector $\hat{H}_{LM} |\psi_S \rangle = \sum_\mathcal{C} a'_\mathcal{C} | \mathcal{C} \rangle$. 

How $\hat{\bf S}_{i} \cdot \hat{\bf S}_j$ acts on a configuration $\mathcal{C}$ depends on the spins at the sites $i,j$. For any pair of sites $i \in A$ and $j \in B$, there is a probability $P_{\uparrow \uparrow} = \frac{2 N^\uparrow}{N} ( 1- \frac{2 N^\uparrow }{N})$ that in this configuration $\mathcal{C}$ the state on $i,j$ is $|\uparrow_i \uparrow_j \rangle$. Similarly, we get $P_{\downarrow \downarrow} = P_{\uparrow \uparrow}$,  $P_{\uparrow \downarrow} = ( 1- \frac{2 N^\uparrow }{N})^2$ and $P_{\downarrow \uparrow} = ( \frac{2 N^\uparrow}{N} )^2$. 

We can use these probabilities to compute the three contributions to $a'_{\mathcal{C}}$,
The diagonal part of $\hat{H}_{LM}$ ($\frac{1}{N} \sum_{ij} \hat{S}^z_i \hat{S}^z_j$) yields when acting on $|\psi_S\rangle$ the following contribution to $a'_{\mathcal{C}}$,
\begin{equation}
		a_{N^\uparrow_B} \frac{1}{4N} (N/2)^2 ( P_{\uparrow \uparrow}+P_{\downarrow \downarrow}-P_{\downarrow \uparrow}-P_{\uparrow \downarrow}).
		\label{eq:Contribution1}
\end{equation}
Here the factor $\frac{1}{4N}$ comes from acting with $\frac{1}{N} \hat{S}^z_i \hat{S}^z_j$, the $(N/2)^2$ is the total number of pairs $i\in A, j \in B$, and $a_{N^\uparrow_B}$ is taken from Eqn.~\eqref{eq:Amplitudes}.

The $\hat{S}^-_i \hat{S}^+_j$ term takes a state from the $(N_B^\uparrow-1)$-sector and brings it into the $N_B^\uparrow$-sector. Therefore, our configuration $\mathcal{C}$ obtains a contribution 
\begin{equation}
	a_{(N^\uparrow_B - 1)} \frac{1}{2N} (N/2)^2 P_{\downarrow \uparrow}.
		\label{eq:Contribution2}
\end{equation}
Similarly, the contribution from the $(N_B^\uparrow+1)$-sector equals 
\begin{equation}
	a_{(N^\uparrow_B + 1)} \frac{1}{2N} (N/2)^2 P_{\uparrow \downarrow}.
	\label{eq:Contribution3}		
\end{equation}
Summing these three contributions Eqs.~\eqref{eq:Contribution1}-\eqref{eq:Contribution3}, and using the identities
\begin{eqnarray}
	\frac{a_{(N^\uparrow_B - 1)}}{a_{N^\uparrow_B }} 
	&=& - \frac{N/2-N^\uparrow_B+1}{N_B^\uparrow} \\
	\frac{a_{(N^\uparrow_B + 1)}}{a_{N^\uparrow_B }} 
	&=& - \frac{N^\uparrow_B+1}{N/2-N_B^\uparrow},
\end{eqnarray}
we find that the amplitude of the configuration $\mathcal{C}$ with given $N^\uparrow_B$ in the vector $\hat{H}_{LM} | \psi_S\rangle$ equals
\begin{equation}
	a'_{\mathcal{C}} = - \frac{1}{4} \left( \frac{N}{4} + 1 \right) a_{N_B^\uparrow}.
\end{equation}
This proves that 
\begin{equation}
	H_{LM} |\psi_S \rangle = - \frac{1}{4} \left( \frac{N}{4} + 1 \right) | \psi_S \rangle,
\end{equation}
and thus that $|\psi_S \rangle$ is the ground state of the Lieb-Mattis Hamiltonian for $S=\frac{1}{2}$.

\subsection{Schwinger boson representation}

The construction for $S=1/2$ in the last paragraph can be extended to general $S$ using the method of Schwinger bosons.\cite{Auerbach:1994vu} In the Schwinger bosons technique, one replaces the spin operators by two sets of bosons,
\begin{eqnarray}
	\hat{S}^+_j & = & \hat{a}^\dagger_j \hat{b}_j \\
	\hat{S}^-_j & = & \hat{b}^\dagger_j \hat{a}_j \\
	\hat{S}^z_j & = & \frac{1}{2} \left( 
		\hat{a}^\dagger_j \hat{a}_j - \hat{b}^\dagger_j \hat{b}_j \right)
\end{eqnarray}
under the constraint that $\hat{a}^\dagger_j \hat{a}_j + \hat{b}^\dagger_j \hat{b}_j = 2S$. A spin coherent state pointing in the ${\bf n}$ direction at site $j$ can be written using Schwinger bosons as
\begin{equation}
	| {\bf n} \rangle_j = \frac{ ( u \hat{a}^\dagger_j + v \hat{b}^\dagger_j )^{2S}}{\sqrt{(2S)!}} |0 \rangle_j
\end{equation}
where $|0 \rangle_j$ is the (unphysical) boson vacuum, $u = e^{i \phi/2} \cos \frac{\theta}{2}$ and $v = e^{-i \phi/2} \sin \frac{\theta}{2}$; compare to Eq.~\eqref{eq:RotatedNeel2}. A state in the opposite direction is expressed as
\begin{equation}
	| -{\bf n} \rangle_j = \frac{ ( i v^* \hat{a}^\dagger_j - i u^* \hat{b}^\dagger_j )^{2S}}{\sqrt{(2S)!}} |0 \rangle_j
\end{equation}
The N\'{e}el state for general $S$ can therefore be written as the product state of an $S_A = NS/2$ spin coherent state on sublattice $A$ and a $S_B=NS/2$ spin coherent on sublattice $B$ in the opposite direction,
\begin{equation}
	|\psi_N ({\bf n}) \rangle = \frac{1}{(NS)!} ( u \hat{a}^\dagger_A + v \hat{b}^\dagger_A )^{NS} ( i v^* \hat{a}^\dagger_B - i u^* \hat{b}^\dagger_B)^{NS} | 0 \rangle
\end{equation}
and the symmetrized state is now
\begin{equation}
	|\psi_S \rangle = \int_{| {\bf n}| = 1} \frac{ d {\bf n}}{4 \pi}  |\psi_N ({\bf n}) \rangle.
\end{equation}
Explicitly writing out this integral gives us
\begin{widetext}
\begin{equation}
	|\psi_S \rangle =
		\frac{(-i)^{NS}}{4 \pi (NS)!}
		\sum_{k,\ell=0}^{NS}
		\binom{NS}{k} \binom{NS}{\ell}
		\left[ \int_0^{2\pi} d\phi \int_0^\pi d\theta \; \sin \theta \;
		u^k v^{NS-k} (-v^*)^{\ell} (u^*)^{NS -\ell} \right]
		(\hat{a}^\dagger_A)^k (\hat{b}^\dagger_A)^{NS-k}
		(\hat{a}^\dagger_B)^{\ell} (\hat{b}^\dagger_B)^{NS-\ell}
		|0 \rangle
	\label{Eq:FullSchwingerSymm}
\end{equation}
The part in the square brackets can only be nonzero when $k+\ell - NS = 0$ due to the integral over $\phi$. Eliminating $\ell$ yields an integral of the form Eq.~\eqref{Eq:IntIdentity},
\begin{equation}
	\left[ \cdots \right] = 2 \pi \int_0^\pi d\theta \; \sin \theta \;
		|u|^{2k} |v|^{2 (NS - k)}
		= \frac{2}{NS + 1} \frac{1}{\binom{NS}{k}}
\end{equation}
The full expression Eq.~\eqref{Eq:FullSchwingerSymm} now becomes
\begin{eqnarray}
	|\psi_S \rangle &=&
	\frac{(-i)^{NS}}{(NS+1)!}
		\sum_{k=0}^{NS} \binom{NS}{k} (-1)^{NS-k} 
		(\hat{a}^\dagger_A \hat{b}^\dagger_B)^k 
		(\hat{b}^\dagger_A \hat{a}^\dagger_B)^{NS-k} | 0 \rangle
	\\
	&=& \frac{(-i)^{NS}}{(NS+1)!}
		\left(\hat{a}^\dagger_A \hat{b}^\dagger_B- \hat{b}^\dagger_A \hat{a}^\dagger_B \right)^{NS} | 0 \rangle.
\end{eqnarray}
\end{widetext}
This final expression is the spin singlet ground state of the Lieb-Mattis Hamiltonian for general $S$.\footnote{This section is based on private communications with H. Katsura.}


\section{General symmetrized superpositions}
\label{Sec:Generalized}

We showed that a symmetric superposition of N\'{e}el states yields the Lieb-Mattis Hamiltonian ground state. A natural follow-up question is: what happens if one takes a superposition of all possible directions of a general initial state?

To answer this question, we consider a spin state $|\psi_0 \rangle$ that is somehow polarized in the $z$-direction. The classical N\'{e}el state polarized in the $z$-direction is an example of such state, but one may also choose a ferromagnet in the $z$-direction, or any eigenstate of $S^z$ for a system of $N$ spin-$S$ degrees of freedom.

Now in general this initial state $|\psi_0 \rangle$ is a superposition of states with different total spin $s$ and total magnetization $m$,
\begin{equation}
	| \psi_0 \rangle = \sum_{s m} a_{sm} | s, m \rangle.
\end{equation}
The total symmetrized state is thus a superposition of the symmetrized states expanded over this $s,m$ basis,
\begin{equation}
	| \psi_S \rangle = \sum_{sm} a_{sm} | \psi_{S} (s,m) \rangle
\end{equation}
where we implicitly defined
\begin{equation}
	|\psi_S (s,m) \rangle = \int \sin \theta d \theta d\phi \; e^{-i \phi S^z} e^{-i \theta S^y} | s , m \rangle.
\end{equation}
There are four different ways $|\psi_S(s,m)\rangle$ can contribute to $|\psi_S\rangle$:
\begin{enumerate}
	\item The symmetrized version of an initial singlet $| 0, 0 \rangle$ is singlet as well, so $|\psi_S (0,0) \rangle = |0,0\rangle$.
	\item The symmetrized version of an initial state $|s, m \rangle$ with $s>0$ and $m=0$ vanishes, $| \psi_S (s,m=0) \rangle = 0$. 
	\item The symmetrized version of an initial state $|s,m \rangle$ with $s>0$ and $m>0$ for $s-m$ even is proportional to the $m=0$ state, $|\psi_S (s,m) \rangle = |s,0 \rangle$.
	\item The symmetrized version of an initial state $| s, m \rangle$ with $s>0$ and $m>0$ for $s-m$ odd vanishes.  
\end{enumerate}

{\em We can thus conclude that the final state $|\psi_S \rangle$ is symmetric if and only if the initial state has a nonzero overlap with a singlet and is otherwise made up of states that vanish.}

For example, the N\'{e}el state is a superposition of the singlet state and other states with $s>0$ but $m=0$. Because $|s, m=0 \rangle$ vanishes when averaged over, the final symmetrized state is just the singlet and hence symmetric. Notice that this is conform the notion of Anderson's Tower of States, which expresses in general the symmetry broken state as a superposition of $m=0$ but $s\geq 0$ states.\cite{Anderson:2011vu}

A corollary of the above statement is that symmetrizing {\em twice} always projects the initial state onto its singlet component: the first symmetrization projects $|\psi_0 \rangle$ onto its $m=0$ components, the second symmetrization makes all terms vanish except for $s=0$.

In order to prove the statements mentioned above, we will expand $e^{-i \phi S^z} e^{-i \theta S^y} | s , m \rangle$ in the basis of $|s,m'\rangle$ states,
\begin{widetext}
\begin{eqnarray}
	\langle s,m' | e^{-i \phi S^z} e^{-i \theta S^y} | s , m \rangle &=&
		e^{-i \phi m'} \langle s, m' | e^{-i \theta S^y} | s m \rangle \\
	&=&
		e^{-i \phi m'}  \sum_{x=\mathrm{max}(0, m'-m)}^{\mathrm{min}(s-m, s+m')}
			(-1)^x
			\frac{\sqrt{ (s+m)! (s-m)! (s+m')! (s-m')!}}{(s-m-x)!(s+m'-x)!x!(x+m-m')!}
		\nonumber \\ &&
			\;\;\;\; \phantom{mmmmmmm} \times
			\cos^{2s+m'-m-2x} \frac{\theta}{2} \sin^{2x+m-m'} \frac{\theta}{2}
\end{eqnarray}
The second line is based on Ref.~\cite{Wigner:1959tu}, Eq. (15.27).

We can now write
\begin{eqnarray}
	| \psi_S (s,m) \rangle &=& \sum_{m'=-s}^s |s, m' \rangle 
		\int_0^\pi \sin \theta d\theta \int_0^{2\pi} d\phi \;
			\langle s,m' |e^{-i \phi S^z} e^{-i \theta S^y} | s , m \rangle
\end{eqnarray}
The integral over $\phi$ only yields a nonzero result whenever $m' = 0$ because of the phase factor $e^{-i\phi m'}$. Note that here the explicit definition of the rotation comes into play: if we would use Eq.~\eqref{Eq:WrongDef}, the extra phase factor $e^{i s \phi}$ implies that we always project onto the $|s , m'=s \rangle$ state. For our original definition of Eq.~\eqref{Eq:DirectionalStates}, we conclude that the symmetrized state is always a $m'=0$ state,
\begin{equation}
	| \psi_S (s,m) \rangle = |s, 0 \rangle 
		\int_0^\pi \sin \theta d\theta \;
			\langle s, 0 | e^{-i \theta S^y} | s , m \rangle
\end{equation}
To see whether the prefactor vanishes or not, let us now compute the integral over $\theta$. For this, we use the identity of Eq.~\eqref{Eq:IntIdentity}, so that, implicitly assuming $m \geq 0$,
\begin{eqnarray}
	| \psi_S (s,m) \rangle &=& |s, 0 \rangle 
		\sum_{x = 0}^{s-m} (-1)^x \frac{s! \sqrt{(s+m)!(s-m)!}}{(s-m-x)! (s-x)! x! (x+m)!}
			I(2s-m-2x,2x+m) \\
	&=& |s, 0 \rangle
	\frac{2\sqrt{(s+m)!(s-m)!} }{(s+1)}
	\sum_{x = 0}^{s-m} (-1)^x 
		\frac{(s-x-m/2)! (x+m/2)! }{(s-m-x)! (s-x)! x! (x+m)!}
\end{eqnarray}
Now in a few limiting cases, this equation can be simplified dramatically. First, if our initial state was the $m=0$ state, we find that the sum over $x$ becomes $\sum_{x=0}^s \frac{(-1)^x}{(s-x)! x!} = 0$ for $s>0$. Therefore, the symmetrized state is actually vanishing.

The second limiting case is $m=s$, in which case
\begin{equation}
	|\psi_S (s,s) \rangle = 
		|s,0 \rangle 
		\frac{2\sqrt{(2s)!} }{(s+1)}
		\left( \frac{(s/2)!}{s!} \right)^2
\end{equation}
Thirdly, observe that the sum over $x$ can be changed into a sum over $x' = s - m - x$. This yields an extra factor of $(-1)^{s-m}$, but other than that, the expression is exactly the same in terms of $x$ or $x'$. Therefore, if $s-m$ is odd-integer, the sum vanishes.

The final case, summarized by $0<m<s$ with $s-m$ an even integer, gives rise to a complicated expression,
\begin{equation}
	|\psi_S (s,m) \rangle = 
		|s,0 \rangle 
		\frac{ 2^{s-m} m ( s/2-1)! (s/2)! ( (s-m-1)/2 )! \sqrt{(s+m)! }}
		{\sqrt{\pi} (s+1) s!((s+m)/2)! \sqrt{(s-m)!}}
\end{equation}
\end{widetext}
which is clearly non-vanishing. This concludes the proof of the four statements at the beginning of this section.

\section{Conclusion and outlook}
\label{Sec:Outlook}

In this manuscript we investigated the properties of a symmetric superposition of all possible directional states. In particular, we showed that a superposition of N\'{e}el states equals the exact symmetric ground state of the Lieb-Mattis Hamiltonian.

Many models that exhibit spontaneous symmetry breaking have an exact symmetric ground state at any finite system size. Our result suggests that in such cases, one can express this ground state as an equal-weight superposition of the {\em symmetry-broken ground states}. For example, linear spin wave (LSW) theory provides us an approximation of a symmetry-broken ground state for the Heisenberg antiferromagnet.\cite{Manousakis:1991iw} Taking a superposition of LSW ground states in different directions will approximate the symmetric ground state of the Heisenberg Hamiltonian. This construction can used in studies of low-energy spectra in exact diagonalization, see for example Ref.~\cite{Wietek:2017wd}. We expect the same phenomenology for $XY$ magnets or $U(1)$ superfluids. A possible extension of our work might include $SU(n)$ symmetric systems.

A notable exception appears for so-called `type B' spontaneous symmetry breaking, such as ferromagnets\cite{Watanabe:2012jn,Watanabe:2014bg}. Here the order parameter commutes with the Hamiltonian so that the ground state, even for finite size systems, is not unique. Consequently, there is no `symmetric' ground state and our results show that one cannot make a symmetric state by superposing, for example, ferromagnets in different directions. The same holds for other `type B' systems such as ferrimagnets.\cite{Rademaker:2019fm}

\acknowledgements

We thank Jasper van Wezel and Aron Beekman for collaboration on a related project that inspired this work. We also thank Hal Tasaki and Hosho Katsura for discussions. This work is supported by the Swiss National Science Foundation via an Ambizione grant (L. R.).


\end{document}